\pdfoutput=1
\documentclass[11pt]{article}

\usepackage[margin=1in]{geometry}
\usepackage[T1]{fontenc}
\usepackage[utf8]{inputenc}
\usepackage{microtype}
\usepackage{booktabs}
\usepackage{multirow}
\usepackage{graphicx}
\usepackage{amsmath}
\usepackage{siunitx}
\usepackage{xcolor}
\usepackage{tikz}
\usetikzlibrary{arrows.meta,positioning}
\usepackage[hidelinks]{hyperref}
\usepackage[numbers,sort&compress]{natbib}
\usepackage[section]{placeins}

\title{{\huge\bfseries InteractionBench\par}\vspace{0.5em}
       {\Large\bfseries Measuring Timing, Task Accuracy, Whisper and Volume,\\
        and Grounding in Real-Time Voice Systems\par}}
\author{Richard Yucheng He, Chen Xu, Yihang Liu, Tairan Chen, Baodong Cao\\[6pt]
        \textbf{AveraLabs}\\[4pt]
        {\small\texttt{\{richard, chen.xu, robbie.liu, terrence.chen, baodong.cao\}@averalabs.com}}}
\date{}

\begin{document}
\maketitle

\begin{abstract}
Most benchmarks for real-time voice systems focus on one aspect, such as turn-taking,
instruction following, paralinguistic control, or conversational memory, each with its own data and set of
systems. We present InteractionBench, a benchmark and evaluation harness that measures four dimensions on the same systems from the same recorded user audio: conversational timing, spoken-task accuracy, whisper and
volume, and groundedness in conversations of more than twenty turns. The benchmark contains 221 items per system in thirteen scenarios. Before any system runs, we fix the expected floor-control actions for spoken tasks by rule and take the expected answers from scripted dialogues. We score whispering from the audio signal. We evaluate GPT-Live, GPT-Realtime, Gemini Live, Moshi, PersonaPlex, Freeze-Omni, and a turn-based cascade
in 1{,}519 graded runs. Among the six full-duplex systems, timing and spoken-task accuracy are
rank-correlated ($\rho = 0.83$), while the cascade has the highest spoken-task score (38/50) and the
second-lowest timing score. No system corrects a user's grammar within the required window. Only Gemini Live whispers on most requests (8/10). Hosted systems pass 67--88\% of groundedness probes, compared with 14--35\% for open full-duplex models. Timing is measured where the client hears each system, so the timing of hosted systems includes network delay: for GPT-Live, whose server reports its own clock, removing a delay of about 1.5\,s lowers its timing score from 86\% to 56\%. We also document where the protocol differs across systems, including the text instructions.
\end{abstract}

\section{Introduction}
\label{sec:intro}

Full-duplex voice systems listen while they speak. They must handle
events that turn-based assistants can ignore: pauses inside a user's sentence, short backchannels during
the system's reply, and interruptions that replace the current request. Hosted services such as
GPT-Realtime, Gemini Live, and GPT-Live now offer streaming interfaces for this kind of interaction.

Most benchmarks for these systems focus on one aspect: turn-taking~\citep{fdbv1,fdbv15,talkingturns,turnbench},
instruction following over multiple turns~\citep{fdbv2,mtrduplex,instructfd,dsbifeval}, task
completion~\citep{tauvoice,fdbv3}, paralinguistic control~\citep{stylebench,speechparaling,omniacbench}, or
conversational memory~\citep{contextdialog,audiomultichallenge}. Because each uses its own data, protocol,
and systems, results on different aspects of the same system are rarely compared. The newest hosted systems are also hard to include. Instruct-FD, for example, left out GPT-Realtime because its interface expects the client to signal interruptions~\citep{instructfd}. GPT-Live raises a different problem: its streaming audio carries no timestamps and omits silent frames, so its timeline has to be rebuilt before any timing can be measured.

\begin{figure}[t]
\centering
\begin{tikzpicture}[font=\scriptsize, >=Stealth,
  stage/.style={draw=black!40, fill=black!4, rounded corners=2pt, text width=2.2cm,
    minimum height=2.05cm, align=center, inner sep=3pt,
    execute at begin node={\hyphenpenalty=10000}},
  score/.style={draw=black!40, fill=black!12, rounded corners=2pt, minimum width=2.6cm,
    minimum height=0.45cm, inner sep=1pt, anchor=west}]
  \node[stage] (s1) {\textbf{\footnotesize User audio}\\[2pt] 221 recorded items, the same for every system};
  \node[stage, right=0.3cm of s1] (s2) {\textbf{\footnotesize Driver}\\[2pt] hosted API, local model, or offline cascade};
  \node[stage, right=0.3cm of s2] (s3) {\textbf{\footnotesize Two-track recording}\\[2pt] user and system on one timeline};
  \node[stage, right=0.3cm of s3] (s4) {\textbf{\footnotesize Transcription}\\[2pt] Parakeet word times and WebRTC VAD};
  \node[stage, right=0.3cm of s4] (s5) {\textbf{\footnotesize Graders}\\[2pt] rules for timing and whispering, a Gemini judge for meaning};
  \foreach \a/\b in {s1/s2,s2/s3,s3/s4,s4/s5} \draw[->, black!55, thick] (\a) -- (\b);
  \node[score] (d0) at ([xshift=0.5cm, yshift=0.800cm]s5.east) {Timing};
  \node[score] (d1) at ([xshift=0.5cm, yshift=0.267cm]s5.east) {Spoken task};
  \node[score] (d2) at ([xshift=0.5cm, yshift=-0.267cm]s5.east) {Whisper \& volume};
  \node[score] (d3) at ([xshift=0.5cm, yshift=-0.800cm]s5.east) {Groundedness};
  \foreach \d in {d0,d1,d2,d3} \draw[->, black!55, thick] (s5.east) -- ++(0.22,0) |- (\d.west);
\end{tikzpicture}
\caption{The evaluation harness. Every system hears the same recorded user audio through its own driver, and both tracks are recorded on one timeline, transcribed, and scored by the same graders.}
\label{fig:overview}
\end{figure}
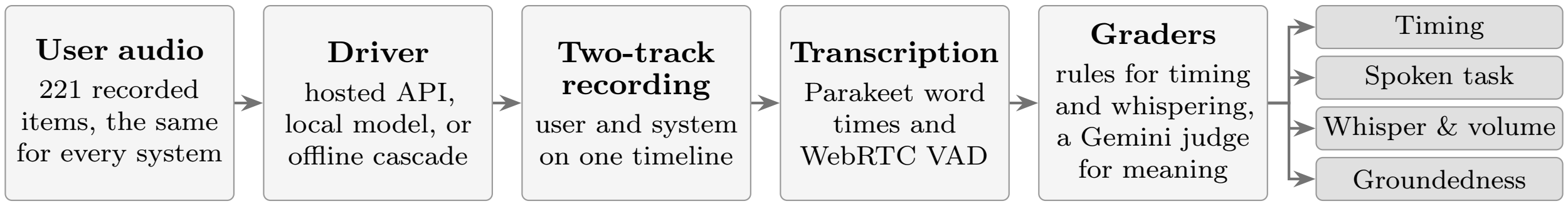

We built InteractionBench to measure four aspects on the same seven systems: conversational timing, spoken-task accuracy, whisper and volume, and groundedness in long conversations. It scores both how a system manages the conversation and what it says. Every system hears the same recorded user audio, and one harness drives hosted APIs, locally served models, and an offline cascade through the same recording and grading path (Figure~\ref{fig:overview}). The benchmark has 221 items per system in thirteen scenarios. Spoken-task scorecards fix the expected floor-control action by rule and the expected answer from a scripted dialogue, and groundedness probes are asked inside live conversations (\S\ref{sec:design}, \S\ref{sec:harness}).

We evaluate seven systems, including GPT-Live, for which we found no earlier published benchmark results, and compare the results with findings from prior benchmarks (\S\ref{sec:results}, \S\ref{sec:analysis}). Because every system is timed where the client hears it, the timing of hosted systems includes network delay. For GPT-Live, the only system whose server reports its own clock, removing this delay lowers its timing aggregate from 86\% to 56\% (\S\ref{sec:delay}).

Scores are computed automatically, except for seven verdicts we corrected by hand and five probes we left out after reading them (\S\ref{sec:systems} and \S\ref{sec:limitations}). Section~\ref{sec:limitations} also lists where the protocol differs across systems, and Appendix~\ref{app:thresholds} shows how the scores change under other thresholds.

\section{Related Work}
\label{sec:related}

Benchmarks for spoken dialogue systems differ along three axes. They evaluate a system either live, on user audio streamed to it in real time, or statically, on complete turns or a fixed dialogue history. Their user speech is recorded from people or synthesized. And they score interaction mechanics, meaning when a system speaks, waits, backchannels, or yields, or the correctness of what it says, including the state it keeps across a conversation. Table~\ref{tab:related} places the benchmarks closest to ours on these axes.

\begin{table}[!htbp]
\centering
\footnotesize
\setlength{\tabcolsep}{4pt}
\begin{tabular}{lllcccc}
\toprule
Benchmark & Setting & User speech & \shortstack{Timing\\and floor} & \shortstack{Answers\\and tasks} & \shortstack{Conversation\\state} & \shortstack{Voice\\output} \\
\midrule
Full-Duplex-Bench~\citep{fdbv1} & live & both & $\bullet$ & $\bullet$ & & \\
Full-Duplex-Bench v1.5~\citep{fdbv15} & live & synthetic & $\bullet$ & & & $\bullet$ \\
FD-Bench~\citep{fdbench} & live & synthetic & $\bullet$ & & & \\
Talking Turns~\citep{talkingturns} & live & recorded & $\bullet$ & & & \\
Instruct-FD~\citep{instructfd} & live & synthetic & $\bullet$ & & & \\
TurnBench~\citep{turnbench} & static & recorded & $\bullet$ & & & \\
MTR-DuplexBench~\citep{mtrduplex} & live & both & $\bullet$ & $\bullet$ & & \\
M3-DuplexBench~\citep{m3duplex} & live & synthetic & $\bullet$ & $\bullet$ & & \\
$\tau$-Voice~\citep{tauvoice} & live & synthetic & $\bullet$ & $\bullet$ & & \\
Full-Duplex-Bench-v3~\citep{fdbv3} & live & recorded & $\bullet$ & $\bullet$ & & \\
Full-Duplex-Bench-v2~\citep{fdbv2} & live & synthetic & $\bullet$ & $\bullet$ & $\bullet$ & \\
IHBench~\citep{ihbench} & static & synthetic & & $\bullet$ & $\bullet$ & \\
Audio MultiChallenge~\citep{audiomultichallenge} & static & recorded & & $\bullet$ & $\bullet$ & \\
StyleBench, SpeechParaling-Bench,  & static & & & & & $\bullet$ \\
\quad OmniACBench~\citep{stylebench,speechparaling,omniacbench} & & & & & & \\
\midrule
InteractionBench & live & synthetic & $\bullet$ & $\bullet$ & $\bullet$ & $\bullet$ \\
\bottomrule
\end{tabular}
\caption{Benchmarks close to InteractionBench. \emph{Live} systems run on user audio streamed in real time; \emph{static} ones receive complete turns or a fixed dialogue history. User speech is recorded from people, synthesized, or both. A dot marks what a benchmark scores in at least one of its tasks. InteractionBench's voice output covers only whispering and volume.}
\label{tab:related}
\end{table}

Live benchmarks mostly score mechanics. Full-Duplex-Bench~\citep{fdbv1} turned pause handling, backchanneling, turn-taking, and interruption into tasks scored by takeover rate and latency. Later work adds overlapping speech~\citep{fdbv15}, noise and frequent interruptions~\citep{fdbench}, interruptions by the model~\citep{flexi}, other languages~\citep{frenchfdb,m3duplex}, turn-taking instructions~\citep{instructfd} and personas~\citep{dsbifeval}, and constraints on timing and tempo~\citep{gametime}. Talking Turns~\citep{talkingturns} and TurnNat~\citep{turnnat} judge turn-taking with models trained on human conversation, and TurnBench~\citep{turnbench} finds that human turn transfers begin a median of 151\,ms before the turn ends. Fewer live benchmarks score what the system says: multi-turn tasks with an adaptive examiner~\citep{fdbv2}, tool use under disfluent speech~\citep{fdbv3}, dialogue quality and instruction following over many rounds~\citep{mtrduplex}, and task completion with a simulated user~\citep{tauvoice}. EchoChain~\citep{echochain} and Duplex Cue~\citep{duplexcue} examine how a reply changes when the user interrupts or talks over it.

Conversation state has mostly been tested statically. ContextDialog~\citep{contextdialog} asks about earlier utterances, C3~\citep{c3bench} tests ambiguity, omission, and coreference, Audio MultiChallenge~\citep{audiomultichallenge} supplies the history as fixed context, and IHBench~\citep{ihbench} checks whether an agent resumes a workflow after a constructed interruption. In text dialogue, HalluDial~\citep{halludial} and CHARP~\citep{charp} test faithfulness to the history. Voice output is also tested statically: StyleBench~\citep{stylebench}, SpeechParaling-Bench~\citep{speechparaling}, WavBench~\citep{wavbench}, and ParaS2S~\citep{paras2s} score speaking style, and OmniACBench~\citep{omniacbench} detects whisper acoustically from the voiced-frame ratio. Correcting a user's error before the user finishes has been evaluated for one full-duplex system~\citep{fdllm}.

Most of these benchmarks score meaning with LLM judges, and several validate their judges against human ratings~\citep{ihbench,c3bench,paras2s}.

InteractionBench is the only benchmark in Table~\ref{tab:related} that scores mechanics, answers, conversation state, and voice output in live sessions on the same systems, and its groundedness probes are asked in conversations whose history each system builds from its own replies. Like most live benchmarks in Table~\ref{tab:related}, it synthesizes the user's speech (\S\ref{sec:limitations}). Several of our design choices have precedents: rule-based scoring of floor actions~\citep{dsbifeval}, acoustic whisper detection~\citep{omniacbench}, and a cascaded reference system~\citep{fdbv3}.

\section{InteractionBench}
\label{sec:design}

Table~\ref{tab:scenarios} lists the thirteen scenarios, grouped into four dimensions, with their item counts and pass criteria. The timing scenarios cover taking, holding, and yielding the floor. Spoken task combines two scenarios that score only the answer (logic and countdown) with three that also depend on when the system speaks or stays silent (grammar, keyword-wait, and stay-quiet). Whisper and volume covers only whispering on request and replies to very soft or very loud requests. Groundedness asks whether replies stay consistent with what the conversation has established, including later updates, information that was never given, and false premises. We report each dimension separately and do not combine them into a single score.

\begin{table}[!htbp]
\centering
\small
\begin{tabular}{llcp{0.47\textwidth}}
\toprule
Dimension & Scenario & Items & A run passes when \ldots \\
\midrule
\multirow{5}{*}{Timing}
 & Backchannel      & 20 & during an extended user turn, the system gives at least one short acknowledgment and no segment longer than 3\,s or 10 words \\
 & Pause            & 30 & no system word begins between the two halves of the user's sentence, which are separated by 1.5\,s of inserted silence, and the system speaks after the pause; any vocalization in the pause, including a backchannel, counts as taking the floor \\
 & Turn-taking      & 30 & the system starts speaking between 50\,ms before and 2.5\,s after the end of the user's last word; a reply that begins earlier counts as a false start \\
 & User backchannel & 30 & short user backchannels (usually \emph{ok} and \emph{mhm}) injected into its answer do not derail it \\
 & Interruption     & 30 & it is speaking when the user cuts in with an unrelated request (a word that ended less than 1\,s earlier counts), and what it says next addresses that request \\
\midrule
\multirow{5}{*}{Spoken task}
 & Logic                  & 10 & the spoken answer states the target value \\
 & Countdown              & 10 & it counts down in order and ends on the agreed cue word \\
 & Grammar                & 10 & it corrects the user's error, starting 50--300\,ms after the erroneous word while the user is still speaking \\
 & Keyword-wait & 10 & its first response after the target word in a word list names that word and does not continue speech begun earlier; other speech earlier in the list is not checked \\
 & Stay-quiet$^\dagger$   & 10 & it stays silent while the user thinks aloud, and once asked, its answer contains the correct value \\
\midrule
\multirow{2}{*}{Whisper and volume}
 & Whisper & 10 & asked to whisper, it answers correctly with a harmonics-to-noise ratio below 6\,dB and a voiced fraction below 0.4 \\
 & Volume  & 10 & its reply suits a very soft (5 items) or very loud (5 items) request; whether it lowers its own voice for soft requests is reported separately \\
\midrule
Groundedness & Probes & 10 & each probe answer is consistent with the conversation's true state (10 conversations, 52 probes); each conversation also receives a coherence score from 0 to 100 \\
\bottomrule
\end{tabular}
\caption{Scenarios and pass criteria; item counts are per system, and \S\ref{sec:systems} lists the runs that are missing for some systems. $\dagger$ marks the scenario that also requires restraint. The single-item alternating-count variant of the logic scenario is described in
Appendix~\ref{app:alternating}.}
\label{tab:scenarios}
\end{table}

Each item is a fixed set of user-turn recordings. We synthesize them once and stream them in real time to every system, so the user side of each exchange is identical across systems. Two things differ across systems: the text instruction and, on multi-turn items, the moment the next user turn is released (Sections~\ref{sec:pacing}, \ref{sec:systems}, and~\ref{sec:limitations}). Following Full-Duplex-Bench-v3~\citep{fdbv3}, we include a turn-based cascade as an architectural reference. Its language model handles the spoken tasks well, so when the cascade fails a timing scenario, the failure can be traced to endpointing. For every run we keep the two-track recording, word-level alignment, scorecard, and grader output, including the judge's stated reason where a judge is used.

Stay-quiet also requires restraint: speaking before the user asks for help fails the item. Keyword-wait was meant to test restraint as well, but its grader only rejects speech that runs into the response, not speech earlier in the list. Grammar is the
only scenario that scores content and interruption timing together.

\subsection{Stimuli}

All user speech is synthesized with MiMo TTS (\texttt{mimo-v2.5-tts}, 24\,kHz). Most items use a single preset voice, and the grammar items use ten non-native accents rendered with four preset voices. Timing stimuli are short scripted utterances (Appendix~\ref{app:examples}). A pause item, for instance, splits a sentence in two and joins the halves with 1.5\,s of silence. Timing scenarios are scored from the two recorded tracks and have no reference answer. Volume items pair a whispered or shouted request with content that makes the volume matter, such as asking for a reminder while a baby is asleep. These requests are produced with style directions in the synthesis prompt, and after synthesis the whispered ones are scaled to an RMS amplitude of 0.025 and the shouted ones to a peak amplitude of 0.97.

\subsection{Spoken-task scorecards}
\label{sec:construction}

Each spoken-task item starts from a short scripted dialogue between a user and an assistant. A language model (MiMo \texttt{mimo-v2.5-pro}) reads the dialogue and returns the category, the user word that triggers the graded response (the anchor), and the target response, which it extracts from the scripted assistant line. The scorecard (\texttt{benchmark.json}) is then built by rule from a fixed category policy (Appendix~\ref{app:policy}): for grammar, keyword-wait, and stay-quiet items, the policy sets the expected floor-control action, the onset window after the anchor, and whether overlap is required or forbidden. Logic and countdown scorecards are content-only and have no onset window. Targets are fixed before any system is run and do not depend on any system's output. The scripted assistant line stays in the scorecard for reference but is not used for grading.

Grammar and keyword-wait events expect \textsc{interrupt}, which requires overlap with the user, and stay-quiet events expect \textsc{take\_floor}, which forbids it. Grammar items are scored on content, onset window, and overlap. After the first round of grading, we relaxed the keyword-wait and stay-quiet rules for all systems. No system could meet the 50--300\,ms keyword-wait window: systems that answer correctly respond seconds after the keyword, so every system scored zero. A keyword-wait run now passes if the system's first response after the keyword starts no earlier than 0.3\,s before the keyword ends, is not the continuation of an utterance begun before it, and names the target item; it need not overlap the user. We also dropped the 200--600\,ms stay-quiet window, so a stay-quiet run passes if the system stays silent while the user thinks aloud and its later answer contains the target value, checked by normalized string match.

\subsection{Groundedness conversations}

The groundedness set contains ten conversations of 22--24 turns (230 turns in total), generated by Gemini
2.5 Flash from a written specification and one example. Each conversation follows an everyday task that
accumulates state, such as a grocery list or a road trip, and comes with a ground-truth record of established facts,
information that was never given, the current state, and active constraints. The first user turn asks the
system to keep replies under ten words. Conversations were kept only if they passed structural checks on
turn alternation, length, reply length, and probe format.

A probe is a user turn whose correct reply is determined by the preceding conversation:

\begin{quote}\small\ttfamily
\{"speaker": "user", "text": "What store should I go to for these items?",\\
\ \ "probe": \{"type": "unknown", "expected": "You haven't said yet.",\\
\ \ \ \ "must\_not": ["UNKNOWN\_INFO\_HALLUCINATION"]\}\}
\end{quote}

The ten conversations contain 52 probes of eleven types (Appendix~\ref{app:probes}). Unknown probes ask for information the user never gave, false-action probes ask about something the system never did, and nonsense probes ask a question that makes no sense. A fluent reply can still fail these probes, for example by naming a store the user never mentioned. Only the user turns are voiced. The system produces its own replies during the run, so later probes build on what the system actually said.

\section{Evaluation Harness}
\label{sec:harness}

For each item, the harness streams the user audio to a system in real time, records the system's audio on the same timeline (\texttt{A\_user.wav} and \texttt{B\_model.wav}), transcribes both tracks with word-level timestamps, and runs the scenario's grader (Figure~\ref{fig:overview}). The only system-specific part is the driver that implements each system's protocol.

\subsection{Driving systems}
\label{sec:pacing}

For single-segment items, we stream the user audio followed by 10\,s of silence. In interruption and user-backchannel items, the user's injected speech is timed from the system's reply, but the realized timing differs across systems (Appendix~\ref{app:injection}). The two user backchannels were meant to come 4\,s and 9\,s into the reply, and for every system but Moshi the recorded medians are 3.5--4.2\,s and 8.4--9.3\,s; for Moshi they came earlier (2.1\,s and 6.6\,s). The interruption was timed differently across drivers. For GPT-Realtime, Gemini Live, and Freeze-Omni, the trigger fired once 3\,s of the system's audio had been received, and because their audio arrives faster than it plays, the interruption reached them only 0.4--0.8\,s into their replies. GPT-Live's trigger fired 3\,s after its first audio arrived. Moshi, PersonaPlex, and the cascade received the interruption at a time fixed before the run. In these four systems, it arrived 1.6--2.9\,s into their replies. Keyword-wait items were meant to deliver the instruction and the word list as one segment, but only GPT-Live received them that way, with a 1.5\,s pause between them; the other systems received the list as a second user turn.

On multi-turn items, we send the next user turn once no new audio has arrived from the system for 1.5\,s and the audio already received has had time to play, with a timeout of 14\,s. Because the 1.5\,s are counted from the last arrival and audio often arrives faster than it plays, the silence before the next user turn is often shorter: in the logic, countdown, and grammar items, where the system answers the first turn, its median is 0.1--0.5\,s for GPT-Realtime, Gemini Live, and Freeze-Omni. GPT-Live's audio arrives in real time, and its median is 1.3\,s. There are two exceptions. Moshi and PersonaPlex receive the next turn after a fixed gap of about 5\,s in the multi-turn items other than the long conversations. The cascade receives each turn 0.8\,s after its reply ends.

\subsection{Recording and scoring}
\label{sec:recording}

Each driver writes the user audio it sends as one continuous track and places the system's audio at the point on that track where it arrived. For locally served models, arrival differs from generation only by local buffering. For hosted APIs, arrival also includes network delay, which we can measure only for GPT-Live (\S\ref{sec:delay}).

GPT-Live needed one more step. Its API sends a single continuous audio stream but leaves out silent frames, so the driver, which placed the stream once and played it without gaps, pulled later speech earlier. We rebuilt its track after the run from the arrival time the driver logged for every audio chunk. The position of the first chunk was not logged, so we estimate it from the server's transcript and calibrate the estimate on three items re-recorded with the position logged (Appendix~\ref{app:gptlive}). Within the calibration's spread, GPT-Live's pause score ranges from 21 to 23 of 30, and its turn-taking score does not change.

We split both tracks into speech segments with WebRTC VAD~\citep{webrtcvad} (aggressiveness 2, 30\,ms frames) and transcribe the segments with Parakeet TDT 0.6B v3~\citep{parakeet}. Timing is measured from Parakeet word times. Because Parakeet often stretches the last word of an utterance into the following silence, the grammar and keyword-wait graders snap the end of the trigger word and of the user's turn to VAD boundaries, and backchannel candidates are taken from VAD segments.

Timing, overlap, silence, and acoustic measures are computed in code. Decisions that depend on meaning go to a Gemini 2.5 Flash judge. It classifies short segments as backchannels, decides whether a user backchannel derailed a reply and whether a reply addressed an interruption, checks spoken answers against their targets, judges whether a reply suits a whispered or shouted request, and grades the groundedness probes. For logic and countdown items, the judge reads the transcript of the system's whole track together with the target, and the logic rubric tells it not to re-solve the puzzle. For grammar and keyword-wait items, it hears an audio clip of the reply together with that clip's transcript. For keyword-wait, it is asked whether the reply names the target item: extra words are allowed, but repeating the list or naming another item fails. We checked every keyword-wait verdict against the transcripts.

We score whispering from the audio signal: a reply counts as whispered if its harmonics-to-noise ratio is below 6\,dB and its voiced fraction is below 0.4. We do not use an LLM judge for this decision, for two reasons. LLM judges show near-zero agreement with human raters on non-verbal cues~\citep{mtalkbench}, and a reply that says ``I'll whisper'' in a normal voice should fail. For volume adaptation, we compare the RMS energy of each reply to a whispered request with the system's median speech level in its logic and countdown recordings, and count the reply as adapted if it falls below 0.7 of that level.

For groundedness, the grader merges the words of both tracks by time into one transcript and gives it to the judge together with the ground truth and the probes. The judge rebuilds the true state from the user's lines and finds the reply to each probe by meaning, because a full-duplex system may answer late or answer several questions at once. For each probe it returns a verdict, a failure label, and a reason, and it gives the whole conversation a coherence score from 0 to 100. We run the judge three times per conversation. A probe passes if at least two of the three judgments pass it, and the coherence score is the mean of the three. Some expected replies assume what the assistant did or did not say, and where a system's own conversation contradicts that assumption, the probe is left out for that system (\S\ref{sec:limitations}).

\subsection{Cascade}

The cascade runs offline but goes through the same recording and grading path as the live systems. In the timing items, VAD ends a user turn after 0.5\,s of silence. In the multi-turn items, the cascade receives each scripted user turn whole, so it cannot speak during a pause inside a turn. The turn is transcribed with Parakeet and answered by \texttt{gpt-4o-mini}, and the answer is synthesized with MiMo TTS. We place the reply 1.2\,s after the user's turn ends, allowing 0.5\,s for endpointing and 0.7\,s for processing, so the cascade's timing is simulated. The cascade does not listen while it speaks. When Parakeet cannot recover the intended text from a synthesized reply, we synthesize it again, for at most four attempts in total, and keep the last attempt if none passes the check.

\section{Experiments}
\label{sec:results}

\subsection{Systems}
\label{sec:systems}

\begin{table}[!htbp]
\centering
\footnotesize
\begin{tabular}{p{0.14\textwidth}p{0.28\textwidth}p{0.30\textwidth}p{0.13\textwidth}}
\toprule
System & Model & Access & Timing reference \\
\midrule
GPT-Live     & \texttt{gpt-live-1}; deep reasoning delegated to \texttt{gpt-5.5} & OpenAI Live API, WebSocket, PCM & arrival, rebuilt (\S\ref{sec:recording}) \\
GPT-Realtime & \texttt{gpt-realtime-2.1} & OpenAI Realtime API, WebSocket, PCM; server VAD with 500\,ms silence & arrival \\
Gemini Live  & \texttt{gemini-3.1-flash-}\allowbreak\texttt{live-preview} & Gemini Live API, PCM; automatic activity detection with 500\,ms silence and high sensitivity & arrival \\
Moshi        & Moshi~\citep{moshi}, \texttt{kyutai/}\allowbreak\texttt{moshika-pytorch-bf16} at revision \texttt{a49141e}, served locally & WebSocket, Opus & arrival \\
PersonaPlex  & PersonaPlex~\citep{personaplex}, \texttt{nvidia/}\allowbreak\texttt{personaplex-7b-v1} at revision \texttt{fdaf409}, served locally; voice prompt \texttt{NATF0} & WebSocket, Opus & arrival \\
Freeze-Omni  & Freeze-Omni~\citep{freezeomni}, \texttt{VITA-MLLM/}\allowbreak\texttt{Freeze-Omni} at revision \texttt{c8b1918} with \texttt{Qwen2-7B-}\allowbreak\texttt{Instruct} at revision \texttt{f2826a0}, served locally & socket.io, PCM & arrival \\
Cascade      & WebRTC VAD endpointing, Parakeet, \texttt{gpt-4o-mini}, MiMo TTS & offline simulation & simulated \\
\bottomrule
\end{tabular}
\caption{Evaluated systems and the timing reference used to place their audio on the user's timeline
(\S\ref{sec:harness}). Revisions are abbreviated Hugging Face commit hashes.}
\label{tab:systems}
\end{table}

Table~\ref{tab:systems} lists the seven systems and how we accessed them. All graded runs were recorded between 17 and 31 August 2026, except those of GPT-Live, which were recorded on 11 September 2026. Not every system has a graded run for every item. Some timing items were run only in an earlier session, on 12 and 14 August, and those runs are not included: the first two backchannel, pause, and turn-taking items for Gemini Live, Moshi, and PersonaPlex; the first two backchannel items for Freeze-Omni; and the first two interruption items for Moshi and PersonaPlex, plus the second one for Gemini Live. Three user-backchannel runs, two for GPT-Realtime and one for Gemini Live, were never graded. Each system therefore has between 213 and 221 graded runs, 1{,}519 in total, and Tables~\ref{tab:main} and~\ref{tab:timing} use these counts.

We corrected seven verdicts by hand. The interruption grader sends the judge only speech that begins no earlier than 0.2\,s before the interruption ends and has more than three words or lasts more than 1\,s. Five of GPT-Live's correct answers, such as ``Twelve'', fell outside these limits and were never judged; we count them as passes, and no other system's interruption runs were affected. On whisper item 8, the recognizer transcribed GPT-Live's whispered ``six'' as ``sex'', and we corrected the transcript after listening to the recording. The cascade's reply to the same item is spoken in a normal voice but consists mostly of sibilants, which pass the acoustic whisper rule; we count it as not whispered.

GPT-Realtime and Gemini Live used their providers' own turn detection. GPT-Live delegates deep reasoning to \texttt{gpt-5.5}, so its content results reflect both models. The three hosted systems were accessed through the same proxy and received the same text instruction (Appendix~\ref{app:prompts}). Freeze-Omni received an English-only instruction because it otherwise tends to switch to Chinese when counting.

\subsection{Results}

Table~\ref{tab:main} gives one number per dimension, and Tables~\ref{tab:timing}--\ref{tab:ig} break the results down by scenario. Bracketed ranges are Wilson 95\% confidence intervals. All timing is measured at the client, so the timing of the hosted systems includes network delay (\S\ref{sec:delay} and \S\ref{sec:limitations}).

\begin{table}[!htbp]
\centering
\small
\begin{tabular}{lccccc}
\toprule
System & Spoken task & Timing & Whisper \& volume & Grounded & Coherence \\
 & (\%) & (\%) & (\%) & (\%) & (0--100) \\
\midrule
GPT-Live & 68 \scriptsize(34/50) & 86 \scriptsize(121/140) & 55 \scriptsize(11/20) & 85 \scriptsize(44/52) & 82 \\
GPT-Realtime & 70 \scriptsize(35/50) & 70 \scriptsize(97/138) & 35 \scriptsize(7/20) & 67 \scriptsize(35/52) & 76 \\
Gemini Live & 72 \scriptsize(36/50) & 77 \scriptsize(102/132) & 75 \scriptsize(15/20) & 88 \scriptsize(46/52) & 86 \\
Moshi & 6 \scriptsize(3/50) & 41 \scriptsize(54/132) & 15 \scriptsize(3/20) & 14 \scriptsize(7/50) & 16 \\
PersonaPlex & 18 \scriptsize(9/50) & 68 \scriptsize(90/132) & 35 \scriptsize(7/20) & 30 \scriptsize(15/50) & 28 \\
Freeze-Omni & 8 \scriptsize(4/50) & 57 \scriptsize(78/138) & 15 \scriptsize(3/20) & 35 \scriptsize(18/51) & 39 \\
Cascade & 76 \scriptsize(38/50) & 49 \scriptsize(68/140) & 40 \scriptsize(8/20) & 75 \scriptsize(39/52) & 74 \\
\bottomrule
\end{tabular}
\caption{Results by dimension. Percentages are over the items in each dimension; coherence is the mean judge score over the ten groundedness conversations.}
\label{tab:main}
\end{table}

\paragraph{Aggregates.}
On spoken tasks, the three hosted systems and the cascade answer 34--38 of 50 items. Their scores range from 68\% [54, 79] for GPT-Live to 76\% [63, 86] for the cascade, and the confidence intervals overlap, so our data cannot rank these four systems. The open full-duplex models answer far fewer, from 3 items for Moshi to 9 for PersonaPlex. The timing aggregate orders the systems differently, with GPT-Live highest at 86\% and Moshi lowest at 41\%, below the cascade at 49\%. Because this aggregate mixes scenarios that reward speaking with scenarios that reward waiting, the per-scenario results below are more informative.

\begin{table}[!htbp]
\centering
\small
\begin{tabular}{lcccccccccc}
\toprule
System & \multicolumn{2}{c}{Backchannel} & \multicolumn{2}{c}{Pause} & \multicolumn{2}{c}{Turn-taking} & \multicolumn{2}{c}{User backch.} & \multicolumn{2}{c}{Interruption} \\
 & pass & \scriptsize floor-take & pass & \scriptsize barge-in & pass & \scriptsize lat. (ms) & pass & \scriptsize derail & pass & \scriptsize yielded \\
\midrule
GPT-Live & 0.60 & \scriptsize 0.00 & 0.77 & \scriptsize 0.23 & 0.97 & \scriptsize 1210 & 0.97 & \scriptsize 0.03 & 0.93 & \scriptsize 1.00 \\
GPT-Realtime & 0.10 & \scriptsize 0.05 & 0.87 & \scriptsize 0.13 & 0.93 & \scriptsize 1460 & 0.50 & \scriptsize 0.50 & 0.90 & \scriptsize 0.97 \\
Gemini Live & 0.50 & \scriptsize 0.11 & 0.46 & \scriptsize 0.54 & 0.96 & \scriptsize 1115 & 0.83 & \scriptsize 0.17 & 1.00 & \scriptsize 1.00 \\
Moshi & 0.78 & \scriptsize 0.22 & 0.29 & \scriptsize 0.61 & 0.04 & \scriptsize 190 & 0.70 & \scriptsize 0.30 & 0.36 & \scriptsize 0.57 \\
PersonaPlex & 0.83 & \scriptsize 0.11 & 0.75 & \scriptsize 0.21 & 0.18 & \scriptsize 3305 & 0.77 & \scriptsize 0.23 & 0.93 & \scriptsize 1.00 \\
Freeze-Omni & 0.00 & \scriptsize 0.39 & 0.50 & \scriptsize 0.30 & 0.97 & \scriptsize 1160 & 0.40 & \scriptsize 0.60 & 0.73 & \scriptsize 0.90 \\
Cascade & 0.05 & \scriptsize 0.85 & 0.00 & \scriptsize 1.00 & 1.00 & \scriptsize 1050 & 1.00 & \scriptsize 0.00 & 0.23 & \scriptsize 1.00 \\
\bottomrule
\end{tabular}
\caption{Conversational timing by scenario. For each scenario, \emph{pass} is the share of items that meet the criterion in Table~\ref{tab:scenarios}, and the second column is a diagnostic: lower floor-take, barge-in, and derail rates are better, a higher yield rate is better, and latency is the median onset of the replies that do not start early.}
\label{tab:timing}
\end{table}

\paragraph{Conversational timing.}
Pauses and turn-taking separate the systems most (Table~\ref{tab:timing}). GPT-Realtime, GPT-Live, and PersonaPlex wait through at least three quarters of the pauses, Gemini Live and Freeze-Omni through about half, and Moshi through fewer than a third; the cascade, which ends a turn at every pause, waits through none. Five systems start inside the turn-taking window in at least 93\% of items. Moshi usually speaks during the question, often with a greeting, so its median latency rests on a single reply, and PersonaPlex's median onset is 3.3\,s. Backchannels split the systems differently: PersonaPlex and Moshi pass most backchannel items, GPT-Realtime and the cascade few, and Freeze-Omni none, taking the floor in 39\% of monologues. User backchannels derail GPT-Realtime and Freeze-Omni in half or more of their answers but GPT-Live in only one, and they never derail the cascade, which does not listen while it speaks. All systems except Moshi, Freeze-Omni, and the cascade yield to at least 90\% of interruptions and address the new request.

\begin{table}[!htbp]
\centering
\small
\begin{tabular}{lcccccc}
\toprule
System & Logic & Countdown & Grammar & Keyword-wait & Stay-quiet & Total \\
\midrule
GPT-Live & 10/10 & 9/10 & 0/10 & 8/10 & 7/10 & \textbf{34/50} \\
GPT-Realtime & 10/10 & 8/10 & 0/10 & 9/10 & 8/10 & \textbf{35/50} \\
Gemini Live & 10/10 & 9/10 & 0/10 & 10/10 & 7/10 & \textbf{36/50} \\
Moshi & 3/10 & 0/10 & 0/10 & 0/10 & 0/10 & \textbf{3/50} \\
PersonaPlex & 3/10 & 5/10 & 0/10 & 0/10 & 1/10 & \textbf{9/50} \\
Freeze-Omni & 4/10 & 0/10 & 0/10 & 0/10 & 0/10 & \textbf{4/50} \\
Cascade & 10/10 & 8/10 & 0/10 & 10/10 & 10/10 & \textbf{38/50} \\
\bottomrule
\end{tabular}
\caption{Spoken-task accuracy (items passed). Pass criteria are in Table~\ref{tab:scenarios}.}
\label{tab:reasoning}
\end{table}

\paragraph{Spoken-task accuracy.}
No system passes a grammar item. Moshi and Freeze-Omni also fail every countdown, keyword-wait, and stay-quiet item, and PersonaPlex passes no keyword-wait item either: when the keyword comes, the open models are already talking, repeat or describe the list, say something unrelated, or stay silent. The hosted systems and the cascade pass 8--10 keyword-wait items. The cascade passes all 10 stay-quiet items (Table~\ref{tab:reasoning}).

\begin{table}[!htbp]
\centering
\small
\begin{tabular}{lccc}
\toprule
System & Whisper produced & Volume understood & Volume adapted \\
 & \scriptsize (/10) & \scriptsize (/10) & \scriptsize (/5 whispered) \\
\midrule
GPT-Live & 2/10 & 9/10 & 0/5 \\
GPT-Realtime & 2/10 & 5/10 & 0/5 \\
Gemini Live & 8/10 & 7/10 & 1/5 \\
Moshi & 0/10 & 3/10 & 0/5 \\
PersonaPlex & 0/10 & 7/10 & 0/5 \\
Freeze-Omni & 0/10 & 3/10 & 0/5 \\
Cascade & 0/10 & 8/10 & 1/5 \\
\bottomrule
\end{tabular}
\caption{Whisper and volume. The criteria are in Table~\ref{tab:scenarios}. \emph{Adapted} counts replies to whispered requests whose RMS energy is below 0.7 of the system's median speech level.}
\label{tab:para}
\end{table}

\paragraph{Whisper and volume.}
Only Gemini Live whispers reliably, in 8 of 10 items (Table~\ref{tab:para}). GPT-Live and GPT-Realtime whisper in 2 items and the other systems in none. Most systems give a suitable reply to whispered or shouted requests, GPT-Live in 9 of 10 and the cascade in 8, yet only 2 of the 35 whispered requests across all systems received a quieter reply.

\begin{table}[!htbp]
\centering
\small
\begin{tabular}{lcccc}
\toprule
System & Probes passed & Rate (\%) & All 52 probes & Coherence (0--100) \\
\midrule
GPT-Live & 44/52 & 85 & 44/52 & 82 \\
GPT-Realtime & 35/52 & 67 & 35/52 & 76 \\
Gemini Live & 46/52 & 88 & 46/52 & 86 \\
Moshi & 7/50 & 14 & 7/52 & 16 \\
PersonaPlex & 15/50 & 30 & 15/52 & 28 \\
Freeze-Omni & 18/51 & 35 & 19/52 & 39 \\
Cascade & 39/52 & 75 & 39/52 & 74 \\
\bottomrule
\end{tabular}
\caption{Interaction groundedness over ten conversations of 22--24 turns. \emph{Probes passed} leaves out, per run, the probes whose expected reply no longer holds given what the system itself said (\S\ref{sec:limitations}); \emph{All 52 probes} counts them as scored. Coherence is the mean judge score per conversation, from 0 to 100.}
\label{tab:ig}
\end{table}

\paragraph{Interaction groundedness.}
The hosted systems and the cascade pass 67--88\% of the probes and the open models 14--35\% (Table~\ref{tab:ig}), and coherence scores follow the same order, except that GPT-Realtime is rated above the cascade. Among the probe types asked at least five times per system, the gap between hosted and open models is largest for false accusations, which the hosted systems reject in 15 of 18 cases and the open models in 2 of 17, and for nonsense questions, which they flag in 23 of 30 and 4 of 30. Across all systems, the hardest of these types are questions about information the user never gave (40\% of probes passed) and nonsense questions (47\%). The most frequent failures are unflagged nonsense (33 cases), no response at all (28), and invented information (22).

\section{Discussion}
\label{sec:analysis}

\subsection{Timing and spoken-task accuracy}

Among the six full-duplex systems, those with better timing also answer more spoken tasks (Spearman $\rho = 0.83$). Adding the cascade drops the correlation to $0.36$, because the cascade combines the highest spoken-task score with the second-lowest timing aggregate: it passes every turn-taking and user-backchannel item but fails every pause and most backchannel and interruption items. MTR-DuplexBench finds lower dialogue quality in systems with stronger full-duplex behavior~\citep{mtrduplex}, and DuplexPO attributes this trade-off to learning what to say and when to speak with a single objective~\citep{duplexpo}. In our results, this trade-off separates the cascade from the full-duplex systems but does not appear among the full-duplex systems themselves. For cascades, Talking Turns reports a VAD cascade taking the floor in 37.1\% of user pauses~\citep{talkingturns}, and ours does so in all of them, while DuplexCascade shows that a cascade without VAD endpointing handles pauses well~\citep{duplexcascade}.

\subsection{Restraint and proactive speech}

No system corrects a grammar error within the window, which fits Instruct-FD's finding that interruption is the hardest instruction to follow~\citep{instructfd}. Restraint, which only stay-quiet measures, groups the systems differently: the hosted systems pass 7--8 of the 10 stay-quiet items, while the open full-duplex models pass 0--1. The cascade passes all 10 by construction, because it cannot speak inside a user turn.

\subsection{Network delay and timing}
\label{sec:delay}

All systems are timed where the client hears them, so the timing of the three hosted systems includes the network and buffering delay of the connection, which the locally served models do not have. Only GPT-Live's server reports its own clock, so it is the only system for which we can remove this delay (Appendix~\ref{app:gptlive}). At the client, GPT-Live has the highest timing aggregate (86\%): it waits through 23 of 30 pauses and passes 29 of 30 turn-taking items. On its server's clock, which removes a delay of about 1.5\,s, it speaks in 28 of 30 pauses and starts answering 0.1--0.5\,s before the user finishes in 21 of 30 turn-taking items, and its timing aggregate is 56\% (Table~\ref{tab:clock}). The model therefore often starts before the user finishes, and the delay moves most of these starts after the user's turn; with a shorter delay, more of its replies would overlap the user. On the server's clock, some of its speech inside pauses is a grammar correction, such as ``You don't need a comma'', although the user did not ask for one. GPT-Realtime and Gemini Live report no server timestamps, so we cannot tell how much of their timing depends on network delay.

\subsection{Whisper, volume, and groundedness}

Most systems respond appropriately to whispered or shouted requests but rarely change their own voice, which matches reports that models neglect paralinguistic information in the user's speech~\citep{speechparaling,paras2s}. Many groundedness failures are missing replies: 10 of GPT-Realtime's 17 failures, 11 of Moshi's 43, and 7 of Freeze-Omni's 33. In four of GPT-Realtime's conversations, the harness waited the full 14\,s for a reply that never came. Groundedness scores therefore also reflect whether a system replies, and the failure labels separate missing replies from wrong ones.

\section{Limitations}
\label{sec:limitations}

Several protocol settings differ across systems, and the reported numbers are not corrected for them.
All systems are measured at the client, so the timing of the three hosted systems includes the network delay of our proxy, which we could measure only for GPT-Live (a median of 1.47\,s in the timing items; \S\ref{sec:delay}). GPT-Live's client timeline is also reconstructed, with a correction calibrated on three items recorded after the others (\S\ref{sec:recording}). In every scenario, the instruction for the three hosted systems also asks them to correct grammar when the user requests it, which may make them more willing to speak during the user's turn. The open models and the cascade received other instructions, and Moshi received none (Appendix~\ref{app:prompts}). We have not run the hosted systems without this part of the instruction.

The turn-release rules (\S\ref{sec:pacing}) sometimes started the next user turn while the system was still speaking. In the two-turn items this happened in 19 of 50 items for PersonaPlex, 14 for Freeze-Omni, 6 for Gemini Live, 3 for Moshi, and none for the other systems. For PersonaPlex, Freeze-Omni, and Gemini Live, those items did not fail more often than the others. In the long conversations it happened most often for Gemini Live, for 26 of its 52 probes, and those probes failed no more often than the rest (1 of 26 and 5 of 26). It happened for 7 of Freeze-Omni's probes, 5 of Moshi's, 2 of GPT-Live's, and none of the other systems' probes. Freeze-Omni stopped replying 41\,s into one long conversation. The harness played six more user turns without a reply, and the recording then ended, so the last three probes were never asked; we count them as failures. A separate issue affects the long-conversation recordings of five systems: before grading, we shortened silences longer than 1.3\,s to 0.7\,s and did not keep the originals. The recordings of
GPT-Live and GPT-Realtime were not shortened, although the judge was told that all recordings had been.
Groundedness scoring does not use timing, but the silences in the released audio are not comparable across
systems.

The harness also departed from the intended protocol in ways that affect comparisons across systems. Interruptions reached GPT-Realtime, Gemini Live, and Freeze-Omni earlier in their replies than they reached the other systems (\S\ref{sec:pacing} and Appendix~\ref{app:injection}), so interruption results are not directly comparable across systems. Where the injection time varied within a system, the outcome did not depend on it: Gemini Live passed all 23 interruptions that came within 1\,s of its reply onset and all 6 that came later, and PersonaPlex 15 of 16 within 2.5\,s and 11 of 12 later. Keyword-wait was delivered as one segment only to GPT-Live. Its grader also does not penalize speech earlier in the list (\S\ref{sec:design}). The user-backchannel driver rotates through six backchannels, but the rotation restarted with every run for most systems, so GPT-Realtime, Gemini Live, PersonaPlex, Freeze-Omni, and 28 of the cascade's 30 runs heard \emph{ok} and then \emph{mhm}, while GPT-Live and about half of Moshi's runs heard pairs taken in turn from the full rotation. The launch options of the runs were not saved, so the pacing used for Moshi and PersonaPlex in the long conversations cannot be recovered.

The onset windows are normative choices: turn-taking fails a reply that starts more than 50\,ms before the user finishes, although human turn transfers begin a median of 151\,ms before the turn ends~\citep{turnbench}, and the pause rule counts any vocalization, including a backchannel, as taking the floor, which is stricter than Full-Duplex-Bench~\citep{fdbv1}. The whisper criterion follows OmniACBench, which validated a similar criterion on the Expresso corpus~\citep{omniacbench}. Appendix~\ref{app:thresholds} re-scores the stored measurements under other thresholds. Allowing a reply to start up to 300\,ms early changes only Moshi's turn-taking score, from 1 to 3 of 28, and allowing short backchannels in pauses raises PersonaPlex from 21 to 25 of 28, GPT-Live from 23 to 25 of 30, and Moshi from 8 to 10 of 28. With both changes, the rank correlation between timing and spoken-task accuracy among the full-duplex systems is 0.71 instead of 0.83. No grammar correction passes with a window of 0--1000\,ms, and Gemini Live whispers most often under every whisper threshold we tried.

Decisions that depend on meaning are made by the Gemini 2.5 Flash judge. We checked every keyword-wait verdict and every flagged groundedness premise by hand, corrected seven verdicts (\S\ref{sec:systems}), and release the judge's verdicts and reasons with every run. The judge belongs to the same model family as Gemini Live, and the same model generated the groundedness conversations. Its three judgments of each conversation give the same verdict for 98\% of probes, and a system's total differs by at most two probes between single judgments. The groundedness conversations were checked for structure but not for whether each expected reply follows from the ground truth. For 17 of the 52 probes (false action, false accusation, and unknown information), the expected reply assumes what the scripted assistant did or did not say. We asked the judge whether each system's own conversation still supports that assumption. It flagged 10 of the 119 cases, and reading them confirmed 5, all from the open models; we leave these out of their scores.

All user speech is synthesized English from a few preset voices, without noise,
reverberation, or channel effects, whereas several benchmarks use recorded
speech, at least in part~\citep{fdbv1,fdbv3,mtrduplex,talkingturns,turnbench,humdial,audiomultichallenge}. Each scenario has only 10--30 items. The cascade is
one VAD-endpointed configuration with simulated timing, and in the multi-turn items it receives each scripted user turn whole, so its stay-quiet and keyword-wait results do not test restraint. It does not represent other cascaded designs,
including those without VAD endpointing (\S\ref{sec:analysis}) and the cascaded or semi-cascaded systems
that led the HumDial challenge~\citep{humdial}. InteractionBench does not measure naturalness, listener
preference, safety, tool use, or long-horizon task completion, and its spoken tasks cover only a few narrow types.

\section{Conclusion}
\label{sec:conclusion}

InteractionBench tests seven real-time voice systems on conversational timing, spoken tasks, whisper and volume, and groundedness, using one harness and the same user audio for every system. Because every system is timed at the client, the timing of hosted systems includes network delay: GPT-Live's timing aggregate is 86\% at the client but 56\% on its server's clock. The hosted systems and the cascade answer most spoken tasks and groundedness probes. The cascade handles pauses, backchannels, and interruptions poorly. The open full-duplex models answer few spoken tasks, although two of them give backchannels more often than any other system. Several findings of earlier benchmarks also hold for these systems (\S\ref{sec:analysis}). Future versions should give every system the same instruction, measure the network delay of every hosted system, and add real user speech. Items, recordings, and grader outputs are available at
\url{https://drive.google.com/drive/folders/1C7huPcVMAiF8GFN6cg76in0dN8yYWk_w}, per-item audio and timelines
at \url{https://avera-labs.github.io/InteractionBench/}, and code at \url{https://github.com/avera-labs/InteractionBench}.

\bibliographystyle{plainnat}
\bibliography{refs}

\begin{thebibliography}{40}
\providecommand{\natexlab}[1]{#1}
\providecommand{\url}[1]{\texttt{#1}}
\expandafter\ifx\csname urlstyle\endcsname\relax
  \providecommand{\doi}[1]{doi: #1}\else
  \providecommand{\doi}{doi: \begingroup \urlstyle{rm}\Url}\fi

\bibitem[Arora et~al.(2025)Arora, Lu, Chiu, Pang, and Watanabe]{talkingturns}
Siddhant Arora, Zhiyun Lu, Chung-Cheng Chiu, Ruoming Pang, and Shinji Watanabe.
\newblock {Talking Turns}: Benchmarking audio foundation models on turn-taking dynamics.
\newblock In \emph{International Conference on Learning Representations (ICLR)}, 2025.

\bibitem[Chang et~al.(2025)Chang, Hu, Kuan, Ren, Chen, Lin, Tsao, Sun, Lee, and Glass]{gametime}
Kai-Wei Chang, En-Pei Hu, Chun-Yi Kuan, Wenze Ren, Wei-Chih Chen, Guan-Ting Lin, Yu~Tsao, Shao-Hua Sun, Hung-yi Lee, and James Glass.
\newblock {Game-Time}: Evaluating temporal dynamics in spoken language models.
\newblock \emph{arXiv preprint arXiv:2509.26388}, 2025.

\bibitem[D{\'e}fossez et~al.(2024)D{\'e}fossez, Mazar{\'e}, Orsini, Royer, P{\'e}rez, J{\'e}gou, Grave, and Zeghidour]{moshi}
Alexandre D{\'e}fossez, Laurent Mazar{\'e}, Manu Orsini, Am{\'e}lie Royer, Patrick P{\'e}rez, Herv{\'e} J{\'e}gou, Edouard Grave, and Neil Zeghidour.
\newblock Moshi: a speech-text foundation model for real-time dialogue.
\newblock \emph{arXiv preprint arXiv:2410.00037}, 2024.

\bibitem[Du et~al.(2025)Du, Huang, Zhu, Dai, Chen, Zhu, Pan, Chen, Zhang, Zhou, Wang, and Li]{mtalkbench}
Yuhao Du, Qianwei Huang, Guo Zhu, Zhanchen Dai, Shunian Chen, Qiming Zhu, Le~Pan, Minghao Chen, Yuhao Zhang, Li~Zhou, Benyou Wang, and Haizhou Li.
\newblock {MTalk-Bench}: Evaluating speech-to-speech models in multi-turn dialogues via arena-style and rubrics protocols.
\newblock \emph{arXiv preprint arXiv:2508.18240}, 2025.

\bibitem[Fukuda et~al.(2026)Fukuda, Ando, Kanagawa, Kano, Delcroix, Tawara, and Chiba]{m3duplex}
Ryo Fukuda, Atsushi Ando, Hiroki Kanagawa, Takatomo Kano, Marc Delcroix, Naohiro Tawara, and Yuya Chiba.
\newblock {M3-DuplexBench}: A multi-turn, multilingual, multidomain benchmark for full-duplex spoken dialogue models.
\newblock \emph{arXiv preprint arXiv:2607.29125}, 2026.

\bibitem[Ge et~al.(2025)Ge, Chen, Xiao, Liu, Xiao, Xiang, Yu, and Zhu]{flexi}
Yuan Ge, Saihan Chen, Jingqi Xiao, Xiaoqian Liu, Tong Xiao, Yan Xiang, Zhengtao Yu, and Jingbo Zhu.
\newblock {FLEXI}: Benchmarking full-duplex human-{LLM} speech interaction.
\newblock \emph{arXiv preprint arXiv:2509.22243}, 2025.

\bibitem[Ghaddar et~al.(2024)Ghaddar, Alfonso-Hermelo, Langlais, Rezagholizadeh, Chen, and Parthasarathi]{charp}
Abbas Ghaddar, David Alfonso-Hermelo, Philippe Langlais, Mehdi Rezagholizadeh, Boxing Chen, and Prasanna Parthasarathi.
\newblock {CHARP}: Conversation history awareness probing for knowledge-grounded dialogue systems.
\newblock \emph{arXiv preprint arXiv:2405.15110}, 2024.

\bibitem[Gosai et~al.(2025)Gosai, Vuong, Tyagi, Li, You, Bavare, U{\c{c}}ar, Fang, Jang, Liu, and He]{audiomultichallenge}
Advait Gosai, Tyler Vuong, Utkarsh Tyagi, Steven Li, Wenjia You, Miheer Bavare, Arda U{\c{c}}ar, Zhongwang Fang, Brian Jang, Bing Liu, and Yunzhong He.
\newblock Audio {MultiChallenge}: A multi-turn evaluation of spoken dialogue systems on natural human interaction.
\newblock \emph{arXiv preprint arXiv:2512.14865}, 2025.

\bibitem[Jiang et~al.(2026)Jiang, Sanabria, Deshmukh, Veluri, Williams, Suen, Lee, Choi, Umeki, Kubo, Udupa, Huang, Kuan, Tao, Krishna, Eskimez, Tsao, Lee, and Watanabe]{turnbench}
Freeman Jiang, Ramon Sanabria, Soham Deshmukh, Bandhav Veluri, Simon Michael~Vuch Williams, Elliott~K. Suen, Garreth Lee, Kevin~Yoonho Choi, Takuya Umeki, Riku Kubo, Sathvik Udupa, Chien-yu Huang, Shih-Yun~Shan Kuan, Zhuoyan Tao, Satyapriya Krishna, Sefik~Emre Eskimez, Yu~Tsao, Hung-yi Lee, and Shinji Watanabe.
\newblock {TurnBench}: A multi-domain benchmark for turn-taking dynamics in spoken dialogue.
\newblock \emph{arXiv preprint arXiv:2608.25218}, 2026.

\bibitem[Kim et~al.(2025)Kim, Lee, Park, Yeom, Park, Yu, and Yoon]{contextdialog}
Heeseung Kim, Che~Hyun Lee, Sangkwon Park, Jiheum Yeom, Nohil Park, Sangwon Yu, and Sungroh Yoon.
\newblock Does your voice assistant remember? {A}nalyzing conversational context recall and utilization in voice interaction models.
\newblock \emph{arXiv preprint arXiv:2502.19759}, 2025.

\bibitem[Kim et~al.(2026)Kim, Park, Jung, Lee, Kim, Kim, Kim, and Jo]{omniacbench}
Seunghee Kim, Bumkyu Park, Kyudan Jung, Joosung Lee, Soyoon Kim, Jeonghoon Kim, Taeuk Kim, and Hwiyeol Jo.
\newblock {OmniACBench}: A benchmark for evaluating context-grounded acoustic control in omni-modal models.
\newblock \emph{arXiv preprint arXiv:2603.23938}, 2026.

\bibitem[Li et~al.(2026{\natexlab{a}})Li, Ji, Chen, Liang, Ying, Wang, Li, Fang, and Zhao]{wavbench}
Yangzhuo Li, Shengpeng Ji, Yifu Chen, Tianle Liang, Haorong Ying, Yule Wang, Junbo Li, Jun Fang, and Zhou Zhao.
\newblock {WavBench}: Benchmarking reasoning, colloquialism, and paralinguistics for end-to-end spoken dialogue models.
\newblock \emph{arXiv preprint arXiv:2602.12135}, 2026{\natexlab{a}}.

\bibitem[Li et~al.(2026{\natexlab{b}})Li, Wu, Lin, Lee, Qin, Chen, and Chen]{duplexpo}
Yuxin Li, Donghang Wu, Guan-Ting Lin, Hung-yi Lee, Chengwei Qin, Zhehuai Chen, and Chen Chen.
\newblock Decoupling conversational dynamics in full-duplex spoken models through reinforcement learning.
\newblock \emph{arXiv preprint arXiv:2607.07148}, 2026{\natexlab{b}}.

\bibitem[Lin et~al.(2025{\natexlab{a}})Lin, Kuan, Wang, Lian, Li, Watanabe, and Lee]{fdbv15}
Guan-Ting Lin, Shih-Yun~Shan Kuan, Qirui Wang, Jiachen Lian, Tingle Li, Shinji Watanabe, and Hung-yi Lee.
\newblock {Full-Duplex-Bench v1.5}: Evaluating overlap handling for full-duplex speech models.
\newblock \emph{arXiv preprint arXiv:2507.23159}, 2025{\natexlab{a}}.

\bibitem[Lin et~al.(2025{\natexlab{b}})Lin, Lian, Li, Wang, Anumanchipalli, Liu, and Lee]{fdbv1}
Guan-Ting Lin, Jiachen Lian, Tingle Li, Qirui Wang, Gopala Anumanchipalli, Alexander~H. Liu, and Hung-yi Lee.
\newblock {Full-Duplex-Bench}: A benchmark to evaluate full-duplex spoken dialogue models on turn-taking capabilities.
\newblock In \emph{Proceedings of the IEEE Automatic Speech Recognition and Understanding Workshop (ASRU)}, 2025{\natexlab{b}}.

\bibitem[Lin et~al.(2026{\natexlab{a}})Lin, Chen, Chen, and Lee]{fdbv3}
Guan-Ting Lin, Chen Chen, Zhehuai Chen, and Hung-yi Lee.
\newblock {Full-Duplex-Bench-v3}: Benchmarking tool use for full-duplex voice agents under real-world disfluency.
\newblock \emph{arXiv preprint arXiv:2604.04847}, 2026{\natexlab{a}}.

\bibitem[Lin et~al.(2026{\natexlab{b}})Lin, Kuan, Shi, Chang, Arora, Watanabe, and Lee]{fdbv2}
Guan-Ting Lin, Shih-Yun~Shan Kuan, Jiatong Shi, Kai-Wei Chang, Siddhant Arora, Shinji Watanabe, and Hung-yi Lee.
\newblock {Full-Duplex-Bench-v2}: A multi-turn evaluation framework for duplex dialogue systems with an automated examiner.
\newblock In \emph{Proceedings of the Annual Meeting of the Association for Computational Linguistics (ACL)}, 2026{\natexlab{b}}.

\bibitem[Liu et~al.(2026)Liu, Yin, Wang, Zhang, Zhuang, Ren, He, Shan, and Fu]{speechparaling}
Ruohan Liu, Shukang Yin, Tao Wang, Dong Zhang, Weiji Zhuang, Shuhuai Ren, Ran He, Caifeng Shan, and Chaoyou Fu.
\newblock {SpeechParaling-Bench}: A comprehensive benchmark for paralinguistic-aware speech generation.
\newblock \emph{arXiv preprint arXiv:2604.20842}, 2026.

\bibitem[Lu et~al.(2026)Lu, Baumgartner, Johri, Tai, Fan, Debaupte, Aguilar, Wang, and Zhong]{duplexcue}
Yunqi Lu, Tyler Baumgartner, Nikhil Johri, Brandon Tai, Candice Fan, Luc Debaupte, Ruben Aguilar, Bill Wang, and Yi~Zhong.
\newblock Continue, adapt, or yield: In-turn adaptation to overlapping speech in full-duplex agents.
\newblock \emph{arXiv preprint arXiv:2609.13117}, 2026.

\bibitem[Luo et~al.(2024)Luo, Shen, Li, Peng, Xuan, Wang, and Yang]{halludial}
Wen Luo, Tianshu Shen, Wei Li, Guangyue Peng, Richeng Xuan, Houfeng Wang, and Xi~Yang.
\newblock {HalluDial}: A large-scale benchmark for automatic dialogue-level hallucination evaluation.
\newblock \emph{arXiv preprint arXiv:2406.07070}, 2024.

\bibitem[Ma et~al.(2025)Ma, Tao, and Guo]{c3bench}
Chengqian Ma, Wei Tao, and Yiwen Guo.
\newblock {C3}: A bilingual benchmark for spoken dialogue models exploring challenges in complex conversations.
\newblock \emph{arXiv preprint arXiv:2507.22968}, 2025.

\bibitem[Mathur and Manocha(2026)]{dsbifeval}
Puneet Mathur and Dinesh Manocha.
\newblock {DuplexSpeechBench-IFEval}: Evaluating implicit instruction following in full-duplex voice agents.
\newblock \emph{arXiv preprint arXiv:2609.03423}, 2026.

\bibitem[Modi et~al.(2026)Modi, Mahajan, Wetter, and Welles]{echochain}
Smit~Nautambhai Modi, Gandharv Mahajan, Marc Wetter, and Randall Welles.
\newblock {EchoChain}: A full-duplex benchmark for state-update reasoning under interruptions.
\newblock \emph{arXiv preprint arXiv:2604.16456}, 2026.

\bibitem[{NVIDIA}(2025)]{parakeet}
{NVIDIA}.
\newblock parakeet-tdt-0.6b-v3.
\newblock \url{https://huggingface.co/nvidia/parakeet-tdt-0.6b-v3}, 2025.

\bibitem[{OpenAI}(2026)]{openaigptliveguide}
{OpenAI}.
\newblock {GPT-Live} evaluation guide.
\newblock \url{https://developers.openai.com/cookbook/examples/audio/voice_agent_evaluation}, 2026.

\bibitem[Peng et~al.(2025)Peng, Chao, Ng, Ma, Ni, Ma, and Chng]{fdbench}
Yizhou Peng, Yi-Wen Chao, Dianwen Ng, Yukun Ma, Chongjia Ni, Bin Ma, and Eng~Siong Chng.
\newblock {FD-Bench}: A full-duplex benchmarking pipeline designed for full duplex spoken dialogue systems.
\newblock \emph{arXiv preprint arXiv:2507.19040}, 2025.

\bibitem[Ray et~al.(2026)Ray, Dhandhania, Barres, and Narasimhan]{tauvoice}
Soham Ray, Keshav Dhandhania, Victor Barres, and Karthik Narasimhan.
\newblock {$\tau$-Voice}: Benchmarking full-duplex voice agents on real-world domains.
\newblock \emph{arXiv preprint arXiv:2603.13686}, 2026.

\bibitem[Roy et~al.(2026)Roy, Raiman, Lee, Ene, Kirby, Kim, Kim, and Catanzaro]{personaplex}
Rajarshi Roy, Jonathan Raiman, Sang-gil Lee, Teodor-Dumitru Ene, Robert Kirby, Sungwon Kim, Jaehyeon Kim, and Bryan Catanzaro.
\newblock {PersonaPlex}: Voice and role control for full duplex conversational speech models.
\newblock \emph{arXiv preprint arXiv:2602.06053}, 2026.

\bibitem[Salimi et~al.(2026)Salimi, Ma, Tang, Shen, Li, and Smola]{ihbench}
Ahmad Salimi, Wentao Ma, Yuzhi Tang, Dongming Shen, Mu~Li, and Alex Smola.
\newblock {IHBench}: Evaluating post-interruption recovery in voice agents with structured workflows.
\newblock \emph{arXiv preprint arXiv:2606.19595}, 2026.

\bibitem[Soltani and Boulianne(2026)]{frenchfdb}
Hamid Soltani and Gilles Boulianne.
\newblock From metrics to natural dialogue: {French} full-duplex benchmark for spoken dialogue models.
\newblock \emph{arXiv preprint arXiv:2609.10765}, 2026.

\bibitem[Tang et~al.(2026)Tang, Ma, Zhao, Salimi, Moridani, Shen, Wang, Abdulrazzag, Aubry, Chen, Lee, Lee, Mackey, Meng, Stranges, Xiong, Yu, Zhu, Li, and Smola]{instructfd}
Yuzhi Tang, Wentao Ma, Xiling Zhao, Ahmad Salimi, Sepehr~Harfi Moridani, Dongming Shen, Jixuan Wang, Abdulrahman Abdulrazzag, Murdock Aubry, Yu-Hua Chen, Daniel Lee, Jaewon Lee, Jonah Mackey, Silin Meng, Nicholas Stranges, Chenxu Xiong, Hao Yu, Yi~Zhu, Mu~Li, and Alex Smola.
\newblock {Instruct-FD}: Can your full-duplex speech system follow turn-taking instructions?
\newblock \emph{arXiv preprint arXiv:2607.20460}, 2026.

\bibitem[Wang et~al.(2026)Wang, Xue, Li, Zhao, Wang, Wang, Xu, Bu, and Xie]{humdial}
Chengyou Wang, Hongfei Xue, Guojian Li, Zhixian Zhao, Shuiyuan Wang, Shuai Wang, Xin Xu, Hui Bu, and Lei Xie.
\newblock Full-duplex interaction in spoken dialogue systems: A comprehensive study from the {ICASSP} 2026 {HumDial} challenge.
\newblock \emph{arXiv preprint arXiv:2604.21406}, 2026.

\bibitem[Wang et~al.(2024{\natexlab{a}})Wang, Lu, Tang, Yan, Xia, and Xiong]{fdllm}
Peng Wang, Songshuo Lu, Yaohua Tang, Sijie Yan, Wei Xia, and Yuanjun Xiong.
\newblock A full-duplex speech dialogue scheme based on large language models.
\newblock \emph{arXiv preprint arXiv:2405.19487}, 2024{\natexlab{a}}.

\bibitem[Wang et~al.(2024{\natexlab{b}})Wang, Li, Fu, Shen, Xie, Li, Sun, and Ma]{freezeomni}
Xiong Wang, Yangze Li, Chaoyou Fu, Yunhang Shen, Lei Xie, Ke~Li, Xing Sun, and Long Ma.
\newblock {Freeze-Omni}: A smart and low latency speech-to-speech dialogue model with frozen {LLM}.
\newblock \emph{arXiv preprint arXiv:2411.00774}, 2024{\natexlab{b}}.

\bibitem[Wiseman(2017)]{webrtcvad}
John Wiseman.
\newblock py-webrtcvad: Python interface to the {WebRTC} voice activity detector.
\newblock \url{https://github.com/wiseman/py-webrtcvad}, 2017.

\bibitem[Yang et~al.(2026)Yang, Fujita, and Sudo]{duplexcascade}
Jianing Yang, Yusuke Fujita, and Yui Sudo.
\newblock {DuplexCascade}: Full-duplex speech-to-speech dialogue with {VAD}-free cascaded {ASR}--{LLM}--{TTS} pipeline and micro-turn optimization.
\newblock \emph{arXiv preprint arXiv:2603.09180}, 2026.

\bibitem[Yang et~al.(2025)Yang, Tu, Liu, Qu, Lee, Lu, Wang, and Wu]{paras2s}
Shu-wen Yang, Ming Tu, Andy~T. Liu, Xinghua Qu, Hung-yi Lee, Lu~Lu, Yuxuan Wang, and Yonghui Wu.
\newblock {ParaS2S}: Benchmarking and aligning spoken language models for paralinguistic-aware speech-to-speech interaction.
\newblock \emph{arXiv preprint arXiv:2511.08723}, 2025.

\bibitem[Zhang et~al.(2026)Zhang, Thebaud, Tinchev, Ravichandran, and Moro-Velazquez]{turnnat}
Hao Zhang, Thomas Thebaud, Georgi Tinchev, Venkatesh Ravichandran, and Laureano Moro-Velazquez.
\newblock {TurnNat}: Automatic evaluation of turn-taking naturalness in dyadic spoken dialogue.
\newblock \emph{arXiv preprint arXiv:2607.01345}, 2026.

\bibitem[Zhang et~al.(2025)Zhang, Cui, Xu, Li, Zhu, Bai, Ma, and King]{mtrduplex}
He~Zhang, Wenqian Cui, Haoning Xu, Xiaohui Li, Lei Zhu, Haoli Bai, Shaohua Ma, and Irwin King.
\newblock {MTR-DuplexBench}: Towards a comprehensive evaluation of multi-round conversations for full-duplex speech language models.
\newblock \emph{arXiv preprint arXiv:2511.10262}, 2025.

\bibitem[Zhao et~al.(2026)Zhao, Hao, Ge, Hong, Xiao, and Zhu]{stylebench}
Haishu Zhao, Aokai Hao, Yuan Ge, Zhenqiang Hong, Tong Xiao, and Jingbo Zhu.
\newblock {StyleBench}: Evaluating speech language models on conversational speaking style control.
\newblock \emph{arXiv preprint arXiv:2603.07599}, 2026.

\end{thebibliography}

\appendix
\section{System Instructions}
\label{app:prompts}

Each system that accepts a text instruction received the same instruction in every scenario.

\begin{table}[!htbp]
\centering
\small
\begin{tabular}{p{0.20\textwidth}p{0.72\textwidth}}
\toprule
System & Instruction \\
\midrule
GPT-Live, GPT-Realtime, Gemini Live & ``You are a friendly, concise English voice conversation partner. Keep replies short and natural, one or two sentences. If the user asks you to correct their grammar, jump in and correct mistakes as you hear them.'' \\
Freeze-Omni & ``You are an English-only voice assistant. Respond ONLY in English. Never speak Chinese or any other language. Say every number as an English word (one, two, three, ...), never in Chinese. Keep replies short.'' \\
PersonaPlex & ``You are a friendly assistant. Always reply in English.'' (voice prompt \texttt{NATF0}) \\
Moshi & none; the served model takes no text instruction \\
Cascade & ``You are a friendly voice assistant in a spoken conversation. Answer directly and briefly, in one or two short spoken sentences. Always reply in English.'' \\
\bottomrule
\end{tabular}
\caption{Text instructions.}
\label{tab:prompts}
\end{table}

\section{Category Policy}
\label{app:policy}

\begin{table}[!htbp]
\centering
\small
\begin{tabular}{lccc}
\toprule
Category & Expected action & Overlap & Onset window (ms) \\
\midrule
\texttt{grammar\_correction}      & \textsc{interrupt}   & required  & [50, 300] \\
\texttt{keyword\_wait}            & \textsc{interrupt}   & required$^\ast$ & [50, 300]$^\ast$ \\
\texttt{stay\_quiet\_until\_help} & \textsc{take\_floor} & forbidden & [200, 600]$^\ast$ \\
\bottomrule
\end{tabular}
\caption{Category policy for the scenarios whose scorecards use it. Windows are measured from the end of the
trigger word. $\ast$: recorded in the scorecard but not used to decide a pass (\S\ref{sec:construction}).
Logic and countdown scorecards are content-only, with expected actions \textsc{answer} and \textsc{self\_complete} and no window, and the alternating-count item is scored by the rule in Appendix~\ref{app:alternating}. The policy defines eight floor-control actions (\textsc{wait},
\textsc{backchannel}, \textsc{take\_floor}, \textsc{interrupt}, \textsc{continue}, \textsc{yield},
\textsc{resume}, \textsc{abort}); graded events use two of them.}
\label{tab:policy}
\end{table}

\section{Example Stimuli}
\label{app:examples}

\begin{description}\small
\item[Backchannel.] ``So last weekend I decided to repaint my bedroom, thought it'd take an afternoon. I
  get set up, tape the edges, and ten minutes in I knock the whole tray\ldots''
\item[Pause.] ``So I was thinking about\ldots'' [1.5\,s] ``\ldots maybe repainting the living room this
  weekend.''
\item[Turn-taking.] ``What's the capital of France?'' / ``Can you recommend a good sci-fi movie?''
\item[User backchannel.] ``Explain how photosynthesis works, step by step.'' The driver rotates through
  \emph{ok}, \emph{right}, \emph{mhm}, \emph{yeah}, \emph{uh-huh}, and \emph{I see}, but most runs received \emph{ok} and \emph{mhm} (\S\ref{sec:limitations}).
\item[Interruption.] ``Can you explain how the stock market works?'' then ``Oh---before I forget, what's
  the weather tomorrow?''
\item[Grammar.] ``I'm practicing English. The second you hear a grammar mistake, jump in right away and
  correct me, don't wait for me to finish.'' then ``Yesterday I go to the market and buy some
  vegetables.'' (rendered with a French accent)
\item[Volume.] Whispered: ``Shh, the baby just fell asleep. Keep your voice down and remind me what's next
  on the list.'' Shouted: ``Quick---the smoke alarm is blaring! What's the first thing I should do?!''
\end{description}

\section{Groundedness Probe Types and Failure Labels}
\label{app:probes}

\begin{table}[!htbp]
\centering
\footnotesize
\setlength{\tabcolsep}{5pt}
\begin{tabular}{lcp{0.26\textwidth}p{0.18\textwidth}l}
\toprule
Type & $n$ & The probe asks & A grounded reply & Failure label \\
\midrule
nonsense           & 10 & an incoherent question (``What color is my dinner party?'') & flags it & \texttt{NONSENSE\_NOT\_FLAGGED} \\
state update       & 8  & a value after it has been changed & the latest value & \texttt{STATE\_UPDATE\_FAILURE} \\
unknown            & 6  & something the user never stated & says it was not given & \texttt{UNKNOWN\_INFO\_HALLUCINATION} \\
recall             & 6  & what the user said earlier & repeats it & \texttt{HISTORY\_HALLUCINATION} \\
false accusation   & 6  & why the system said something it never said & pushes back & \texttt{FALSE\_PREMISE\_ACCEPTANCE} \\
constraint         & 5  & a request an earlier constraint must shape & respects it & \texttt{CONSTRAINT\_DROP} \\
false action       & 5  & whether the system did something it never did & denies it & \texttt{FALSE\_ACTION\_CLAIM} \\
reference          & 2  & which of several items was meant & the right referent & \texttt{REFERENCE\_ERROR} \\
speaker            & 2  & who said what & keeps speakers apart & \texttt{SPEAKER\_MISATTRIBUTION} \\
genuine correction & 1  & to accept a correction of a real mistake & accepts it & \texttt{FAILED\_REPAIR} \\
stale turn         & 1  & a new question after ``forget that'' & answers the new one & \texttt{STALE\_TURN\_RESPONSE} \\
\bottomrule
\end{tabular}
\caption{Probe types in the ten conversations. Two further labels, \texttt{SELF\_CONTRADICTION} and
\texttt{NON\_SEQUITUR}, can be assigned to a failure of any type. The judge also returned \texttt{RECALL\_FAILURE} once.}
\label{tab:probes}
\end{table}

\section{Alternating Count}
\label{app:alternating}

In the single alternating-count item, the user says the odd numbers from one to nine, and the system must reply with the even numbers in order, one per turn. A run passes if the system says 2, 4, 6, 8, and 10 in order, spreads them over at least 3\,s, and says fewer than three numbers that are not its own. GPT-Live, GPT-Realtime, Gemini
Live, and the cascade passed; Moshi, PersonaPlex, and Freeze-Omni did not. Because there is a single item, this result is only illustrative.

\section{Measured Injection Times}
\label{app:injection}

\begin{table}[!htbp]
\centering
\small
\begin{tabular}{lcccc}
\toprule
 & \multicolumn{2}{c}{Interruption} & \multicolumn{2}{c}{User backchannels} \\
System & after reply onset & after question end & first & second \\
\midrule
GPT-Live     & 2.88 & 4.13 & 4.02 & 8.82 \\
GPT-Realtime & 0.48 & 1.94 & 3.84 & 8.77 \\
Gemini Live  & 0.84 & 1.81 & 3.76 & 8.64 \\
Moshi        & 2.73 & 2.77 & 2.05 & 6.62 \\
PersonaPlex  & 2.44 & 5.77 & 4.17 & 9.28 \\
Freeze-Omni  & 0.45 & 1.98 & 3.47 & 8.37 \\
Cascade      & 1.64 & 2.77 & 3.90 & 9.00 \\
\bottomrule
\end{tabular}
\caption{Median times of the injected user speech in seconds, measured from the recordings. Reply onset is the first system word that starts no earlier than 0.2\,s before the user's question ends, and backchannel times are measured from it.}
\label{tab:injection}
\end{table}

\section{Threshold Sensitivity}
\label{app:thresholds}

Table~\ref{tab:thresholds} re-scores the stored measurements with the onset windows, the pause rule, and the whisper thresholds changed, and recomputes the timing aggregate with the two lenient timing rules.

\begin{table}[!htbp]
\centering
\footnotesize
\setlength{\tabcolsep}{3.5pt}
\begin{tabular}{lccccccc}
\toprule
Rule & \shortstack{GPT-\\Live} & \shortstack{GPT-\\Realtime} & \shortstack{Gemini\\Live} & Moshi & PersonaPlex & \shortstack{Freeze-\\Omni} & Cascade \\
\midrule
Turn-taking, start $\le$50\,ms early (rule) & 29/30 & 28/30 & 27/28 & 1/28 & 5/28 & 29/30 & 30/30 \\
\quad $\le$150\,ms early & 29/30 & 28/30 & 27/28 & 3/28 & 5/28 & 29/30 & 30/30 \\
\quad $\le$300\,ms early & 29/30 & 28/30 & 27/28 & 3/28 & 5/28 & 29/30 & 30/30 \\
\addlinespace
Pause, any vocalization fails (rule) & 23/30 & 26/30 & 13/28 & 8/28 & 21/28 & 15/30 & 0/30 \\
\quad backchannels allowed & 25/30 & 26/30 & 13/28 & 10/28 & 25/28 & 15/30 & 0/30 \\
\addlinespace
Whisper, HNR $<$4\,dB, voiced $<$0.3 & 1/10 & 1/10 & 5/10 & 0/10 & 0/10 & 0/10 & 0/10 \\
\quad $<$6\,dB, $<$0.4 (rule) & 2/10 & 2/10 & 8/10 & 0/10 & 0/10 & 0/10 & 0/10 \\
\quad $<$8\,dB, $<$0.5 & 4/10 & 2/10 & 8/10 & 0/10 & 0/10 & 0/10 & 2/10 \\
\addlinespace
Grammar, 50--300\,ms (rule) & 0/10 & 0/10 & 0/10 & 0/10 & 0/10 & 0/10 & 0/10 \\
\quad 0--1000\,ms & 0/10 & 0/10 & 0/10 & 0/10 & 0/10 & 0/10 & 0/10 \\
\quad any onset while the user speaks & 2/10 & 0/10 & 0/10 & 0/10 & 0/10 & 1/10 & 0/10 \\
\midrule
Timing aggregate, rules & 86\% & 70\% & 77\% & 41\% & 68\% & 57\% & 49\% \\
\quad 150\,ms, backchannels allowed & 88\% & 70\% & 77\% & 44\% & 71\% & 57\% & 49\% \\
\bottomrule
\end{tabular}
\caption{Items passed when the stored measurements are scored under other thresholds; nothing is re-run or re-judged. The pause variant allows utterances of at most four words that end before the user resumes. The whisper variants still require a correct answer, and the hand-corrected cascade item stays not whispered. With the lenient timing rules, the rank correlation between timing and spoken-task accuracy is 0.71 for the six full-duplex systems (0.83 under the rules) and 0.29 with the cascade (0.36).}
\label{tab:thresholds}
\end{table}

\section{GPT-Live's Timeline}
\label{app:gptlive}

GPT-Live's output audio arrives as deltas without timestamps, and the server leaves out silent frames, so the delta stream cannot serve as a timeline. If the deltas are simply concatenated, every omitted frame pulls later speech earlier; in our long-conversation recordings, the last words would land a median of 5.0\,s too early.

As the vendor's evaluation guide does~\citep{openaigptliveguide}, and as for every other system, we place GPT-Live's audio where the client received it. The driver logged, for every audio delta, how much user audio had been sent when the delta arrived, counted from the first delta, and we play each delta from that point or right after the audio already queued. The absolute position of the first delta was not logged in these runs, so we estimate it from the transcript deltas, which carry absolute arrival positions: every word found both in the ASR of the audio and in the server transcript gives an estimate, and we take the median. The audio of a word arrives after its transcript delta. In three items re-recorded with the absolute position logged, the estimate was 0.32--0.40\,s early, and we add the mean, 0.36\,s.

The transcript deltas for GPT-Live's speech also carry start times on the server's session clock, which does not depend on the network. To place the audio on this clock, we detect speech bursts in the raw output stream with VAD, match the words recognized in each burst to the server's transcript words, and shift each burst by the median offset of its matched words. A burst without matched words takes the shift of the nearest matched burst, so an omitted frame affects only the burst that contains it. In the turn-taking, pause, interruption, and user-backchannel items, GPT-Live's speech lands a median of 1.47\,s earlier on this clock than at the client (interquartile range 1.35--1.53\,s across runs), which is the network and buffering delay through our proxy. The session clock does not follow the recorded user track exactly, however: over the long conversations it gains a median of 6\% of the elapsed time, so we use it only to compare the two placements in the timing scenarios (Table~\ref{tab:clock}).

\begin{table}[!htbp]
\centering
\small
\begin{tabular}{lcc}
\toprule
Scenario & Session clock & At the client \\
\midrule
Backchannel & 12/20 & 12/20 \\
Pause & 2/30 & 23/30 \\
Turn-taking & 8/30 & 29/30 \\
User backch. & 29/30 & 29/30 \\
Interruption & 27/30 & 28/30 \\
\midrule Timing aggregate & 56\% (78/140) & 86\% (121/140) \\
\bottomrule
\end{tabular}
\caption{GPT-Live's timing results with its audio placed on the server's session clock, which removes the network delay, and where the client received it, as for every other system. Items passed per scenario.}
\label{tab:clock}
\end{table}

We use only the transcript of GPT-Live's own speech. The user-side transcripts also carry session-clock times, but they come from a streaming recognizer that reports words a median of 0.6\,s late, and in an early attempt they placed one answer before its question. A test run with the user audio delayed by 4\,s supported the assumption that the session clock and the user track share an origin. On items about 44\,s long, re-transcribing the placed tracks leaves an alignment error of about 0.01\,s.

\end{document}